\documentclass[twocolumn,secnumarabic,amssymb, nobibnotes, aps, prl, noeprint]{revtex4-1}

\usepackage{graphicx}
\usepackage{natbib}
\usepackage{textgreek}

\newcommand{\beginsupplement}{%
	\setcounter{table}{0}
	\renewcommand{\thetable}{S\arabic{table}}
	\setcounter{figure}{0}
	\renewcommand{\thefigure}{S\arabic{figure}}}

\begin{document}
	
	\title{Photo-induced anomalous Hall effect in the type-II Weyl-semimetal WTe$_2$ \newline at room-temperature 
	}
	
	\author{Paul Seifert$^1$, Florian Sigger$^{1,2}$, Jonas Kiemle$^1$, Kenji Watanabe$^3$, Takashi Taniguchi$^3$, Christoph Kastl$^4$, Ursula Wurstbauer$^{1,2}$ and Alexander Holleitner$^{1,2}$
	}
	
	\affiliation{1 Walter Schottky Institut and Physics Department, Technical University of Munich, Am Coulombwall 4a, 85748 Garching, Germany.       
		\\
		2 Nanosystems Initiative Munich (NIM), Schellingstr. 4, 80799 München, Germany. 
		\\
		3 Advanced Materials Laboratory, National Institute for Materials Science, Tsukuba, Ibaraki 305-0044, Japan.
		\\
		4 Molecular Foundry, Lawrence Berkeley National Laboratory, One Cyclotron Road, 94720 Berkeley, California, United States.
	}
	\date{Oktober 1, 2018}
	\begin{abstract}
		Using Hall photovoltage measurements, we demonstrate that an anomalous Hall-voltage can be induced in few layer WTe$_2$ under circularly polarized light illumination. By applying a bias voltage along different crystal axes, we find that the photo-induced anomalous Hall conductivity coincides with a particular crystal axis. Our results are consistent with the underlying Berry-curvature exhibiting a dipolar distribution due to the breaking of crystal inversion symmetry. Using a time-resolved optoelectronic auto-correlation spectroscopy, we find that the decay time of the anomalous Hall voltage exceeds the electron-phonon scattering time by orders of magnitude but is consistent with the comparatively long spin-lifetime of carriers in the momentum-indirect electron and hole pockets in WTe$_2$. Our observation suggests, that a helical modulation of an otherwise isotropic spin-current is the underlying mechanism of the anomalous Hall effect.
		
	\end{abstract}
	
	\maketitle
	\section{Introduction}
	In recent years, materials that exhibit a non-trivial topological band-structure and non-zero Berry-curvature have attracted a lot of attention. These properties render the materials a very promising and robust platform for spintronic applications independent of the exact details of material composition or extrinsic influences such as temperature. The Berry-curvature describes the local self-rotation of a quantum wave-packet and can effectively act as a magnetic field largely impacting the electronic properties of a system \cite{Xiao2010}. The Berry-curvature impacts the Hall-conductivity and spin-Hall conductivity, which is the antisymmetric non-dissipative addition to the Ohmic conductivity in the absence of time-reversal symmetry or crystal inversion symmetry, respectively \cite{Haldane2004}. In this context, WTe$_2$ a layered van-der-Waals material, became the subject of extensive research. Monolayer  WTe$_2$ was demonstrated to host both a topologically non-trivial quantum spin-Hall gap \citep{Fei2017,Tang2017,Wu2018}, as well as a Berry curvature dipole that leads to the so-called circular photogalvanic effect when inversion symmetry is broken via an out-of-plane electric field \citep{Xu2018}. Bulk WTe$_2$ is known as the prototypical type-II Weyl semimetal, a material shown to exhibit chiral Weyl-fermions at the surface that break Lorentz-invariance \cite{Soluyanov2015}. Moreover, WTe$_2$ exhibits the highest values of a non-saturating magnetoresistance ever reported \cite{Ali2014}. In contrast to monolayer WTe$_2$, inversion symmetry is intrinsically broken in few-layered and bulk WTe$_2$ \cite{Brown1966}. This renders few layer WTe$_2$ a promising candidate to host non-trivial spin-transport phenomena, as a non-zero Berry curvature on the non-equilibrium Fermi surface is predicted \cite{Zhang2018b}. The Weyl points, which are the linear band-crossing points, represent monopoles of Berry curvature and come pairwise with opposite chirality \cite{Wan2011}. It is proposed, that under circularly polarized (CP) illumination, a momentum-shift of the Weyl-points gives rise to a photoinduced anomalous Hall effect during optical excitation \cite{Chan2016}.\newline In the present work, we explore the transverse conductivity in few layer WTe$_2$ under electrical current flow using polarization- and time-resolved photoexcitation at room temperature. We detect a finite photoinduced transverse voltage that switches sign with the polarity of the applied longitudinal current as well as with the photonr helicity. The helicity dependent transverse conductivity only occurs when the current is injected orthogonal to the crystal's mirror-plane. This observation suggests the intrinsic breaking of inversion symmetry with a respective dipolar structure of the Berry curvature as the underlying physics \cite{Zhang2018c,Yao2008}. We further reveal a characteristic lifetime of the transverse conductivity of $>100$ ps exceeding the reported carrier-phonon scattering time by orders of magnitude \cite{Dai2015}. Therefore, the lifetime of the observed transverse voltage is associated with the reported spin-lifetime of carriers in the momentum-indirect electron and hole pockets in WTe$_2$ \cite{Wang2018,Dai2015}. Our observation suggests a photon helicity induced anisotropy in the spin-Hall conductivity via the excitation of a net spin polarization.\newpage
	\section{}
	\begin{figure}
		\includegraphics[scale=1]{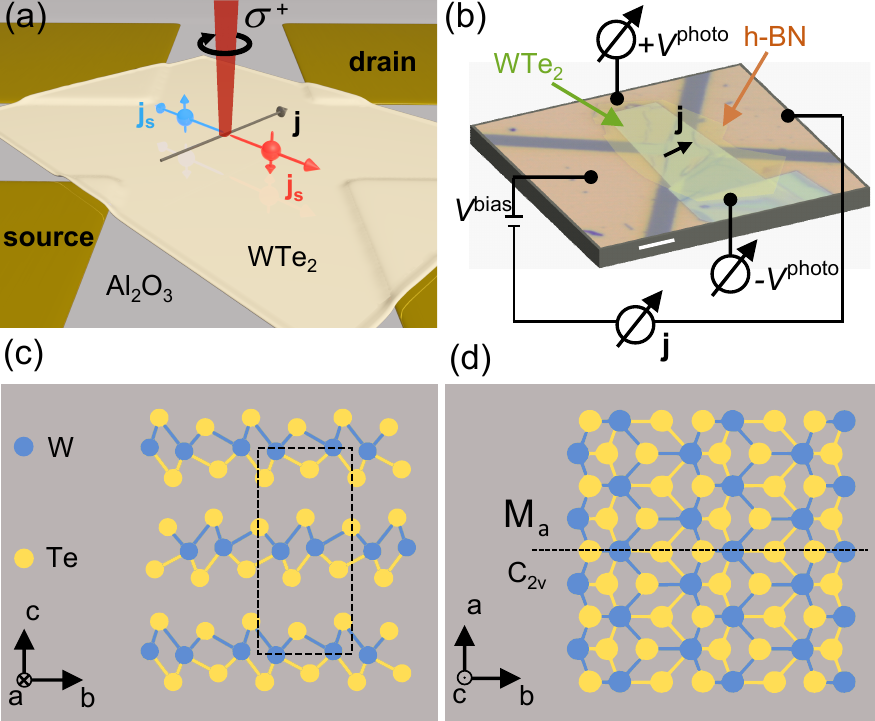}
		\caption{(a) Schematic of an anomalous Hall effect driven by a net spin polarization induced by illumination with circularly polaried light. Red and blue spheres represent opposite spin polarizations flowing perpendicular to a current \textbf{j}. A photo induced, out-of-equillibrium spin polarization leads to a net Hall photovoltage via the spin-Hall effect. (b) Optical microscope image and experimental circuitry of a four-terminal device of WTe$_2$ encapsulated in h-BN. Scale bar is 10 \textmu m. (c) Side view on the crystal structure of few-layer T$_d$-phase WTe$_2$ where a and b (c) denote in-plane (out-of-plane) crystal axis directions. In the T$_d$-phase, inversion symmetry is broken along the b-axis. (d) Top view of the crystal structure of T$_d$-WTe$_2$. The crystal exhibits a mirror plane $\text{M}_{a}$ orthogonal to the a-axis. 
		}
	\end{figure}
	\begin{figure}
		\includegraphics{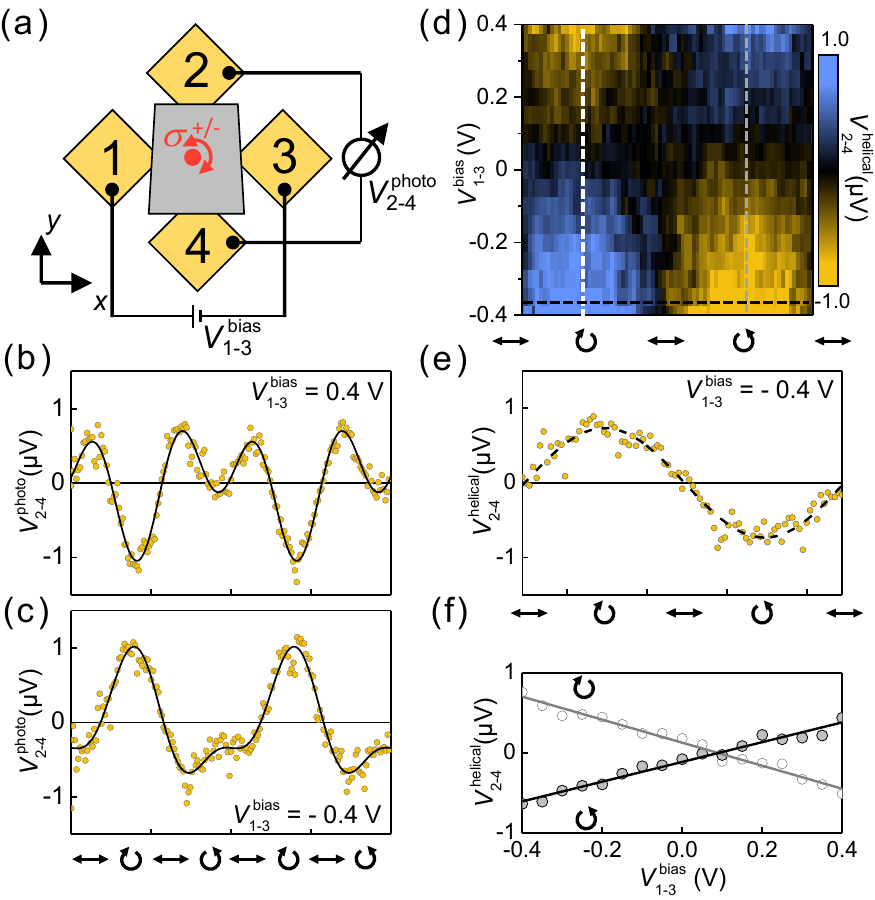}
		\caption{Polarization-resolved transverse photovoltage measurements. (a) Schematic of the measurement geometry. The source-drain bias $V_{1-3}$ is applied between the contacts labeled (1) and (3). The transverse voltage $V_{2-4}$ is measured between the contacts labeled (2) and (4). All measurements are conducted at room temperature. (b) and (c) Polarization dependence of the transverse photovoltage $V^\mathrm{photo}_{2-4}$, for applied bias voltages of $V^\mathrm{bias}_{1-3}$ = 0.4 V and $V^\mathrm{bias}_{1-3}$ = -0.4 V, respectively. The photoresponse is dominated by the photon helicity and the polarization dependence switches polarity with the applied bias $V^\mathrm{bias}_{1-3}$. (d) Bias dependence of the helicity dependent photovoltage $V^\mathrm{helical}_{2-4}$. The modulation with the linear polarization is substracted via a fit procedure (compare supplementary Figs. S1 and S2). (e) $V^\mathrm{helical}_{2-4}$ along the black dashed line in (d). (f) Source-drain bias dependence of $V^\mathrm{helical}_{2-4}$ for circularly right  and left  polarized light, along the grey and white dashed lines in (d).}
	\end{figure}
	We exfoliate individual few-layer flakes of the type-II Weyl-semimetal WTe$_2$ (HQ Graphene) onto a transparent sapphire substrate with pre-fabricated Ti/Au contacts in a four-terminal geometry. We focus a circularly polarized laser with a photon energy of 1.5 eV onto the center of the flake. To avoid degradation effects and to maintain a high device quality, the WTe$_2$-flakes are encapsulated with a thin crystal of high quality hexagonal boron nitride \cite{Ye2016,Watanabe2004}. Figures 1(a) sketches our measurement geometry of the photo-induced anomalous Hall effect. A bias voltage is applied between the source and drain contacts and it drives a current \textbf{j}. A high impedance differential voltage amplifier is wired to the two remainig contacts in perpendicular direction for the measurement of the photo-induced transverse voltage $V^\mathrm{photo}$. Under current flow, a transverse spin-current can be induced for a finite local spin-Hall conductivity $\sigma_H^s$ acting on the non-equilibrium Fermi surface. Without polarized illumination, spin-Hall conductivity $\sigma_H^s$ induces a pure spin-current without any accompanying charge current. However, under circularly polarized illumination the photon's magnetic moment induces a spin polarization and a corresponding net anomalous Hall-conductivity. The latter can be detected as a net Hall voltage orthogonal to the light propagation and the source-drain current density \textbf{j}. Figure 1(b) shows a microscope image of a four-terminal WTe$_2$/h-BN device and the electrical circuitry. The distance between the contacts of 14 \textmu m is chosen to be much larger than the size of our laser spot of $\approx$ 1.5 \textmu m in order to exclude extrinsic effects that stem from an illumination of the metal contacts or crystal boundaries \cite{Xu2018}. Figures 1(c) and (d) sketch the side- and top-view of the layered crystal structure of WTe$_2$. In contrast to most TMDs, which exhibit a hexagonal structure with space-group $D_{6h}$, bulk WTe$_2$ crystallizes in an orthorhombic phase with space group $C_{2v}$ featuring a single mirror plane $\text{M}_{a}$ orthogonal to the a-axis. In the T$_d$-phase, atoms in neighbouring layers are rotated by a spatial angle of $\pi$ with respect to each other, which breaks inversion symmetry along the b-axis and leads to a dipolar structure of the Berry-curvature \cite{Tang2017,Brown1966,Jiang2016,MacNeill2017}. Due to the different dielectric environment above (h-BN) and below (Al$_2$O$_3$) the WTe$_2$ flake, we also cannot exclude a weak breaking of inversion-symmetry along the c-direction of the crystal.

	Figure 2 presents polarization-resolved transverse photovoltage measurements. The laser is focused at the center of the WTe$_2$ flake and a bias voltage $V^\mathrm{bias}_{1-3}$ is applied between source (1) and drain (3), while the photovoltage $V^\mathrm{photo}_{2-4}$ is measured between (2) and (4) (Fig 2 (a)). The laser polarization is modulated with a quarter-wave plate. Figures 2(b) and (c) show the polarization dependent transverse voltage $V^\mathrm{photo}_{2-4}$ for bias voltages of $V^\mathrm{bias}_{1-3}$ = + 0.4 V and $V^\mathrm{bias}_{1-3}$ = - 0.4 V, respectively. A clear transverse $V^\mathrm{photo}_{2-4}$ is detectable and can be modulated with the laser helicity. Figure 2 (d) shows selectively the helicity dependent photovoltage contribution $V^\mathrm{helical}_{2-4}$ as a function of bias voltages $V^\mathrm{bias}_{1-3}$ and laser polarization. The helicity independent contributions are subtracted via a fit procedure according to their characteristic frequency with the angle of the quarter-wave plate (compare supplementary Figs. S1 and S2) \cite{Seifert2018}. Line-cuts through the helical photovoltage map (Fig. 2(d)) are drawn in Fig. 2(e) along the polarization axis and Fig. 2(f) along the bias axis for different laser helicities. The transverse photovoltage $V^\mathrm{helical}_{2-4}$ changes polarity with the laser helicity and depends linearly on the longitudinal bias-voltage $V^\mathrm{bias}_{1-3}$. 
	\newline In order to determine how the transverse photovoltage depends on the crystal axis, we perform measurements on a sample, where the crystal axes are aligned to the measurement axes of our four-terminal contact geometry during sample fabrication. The \textit{x}-axis (\textit{y}-axis, compare Fig. 2 (a)) of our measurement geometry coincides with the b-axis (a-axis) of this WTe$_2$ crystal as determined by polarization resolved Raman spectroscopy (supplementary Fig. S3) \cite{Song2016}.
	\begin{figure}
		\includegraphics{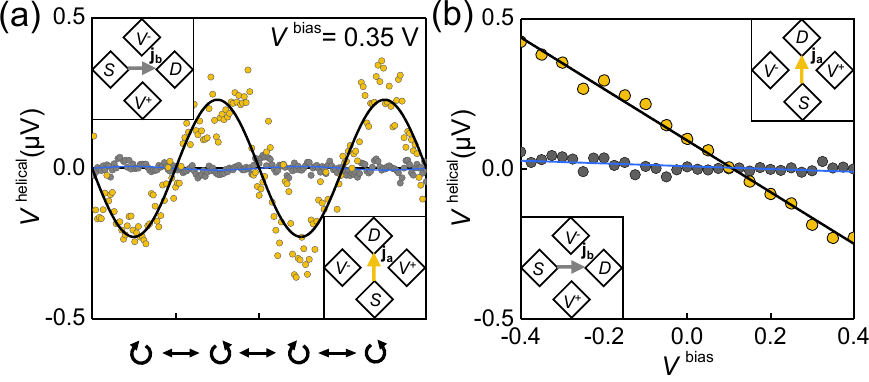}
		\caption{Crystal axis dependent photovoltage. (a) Comparison of the transverse helicity dependent $V^\mathrm{helical}$ induced in the sample for a bias $V^\mathrm{bias}$ applied along the crystal axes orthogonal (yellow) and parallel (grey) to the crystals mirror plane $\text{M}_a$. (b) Amplitude of $V^\mathrm{helical}$ vs. the applied bias $V^\mathrm{bias}$ along the two different crystal axes. Respective measurement configurations are indicated by the insets.}
	\end{figure}
	Figure 3 (a) shows the helicity dependent transverse photovoltage for a bias applied between contacts (1) and (3) (b-axis, grey dots) and between contacts (2) and (4) (a-axis, yellow dots). For a bias applied along the b-axis, no distinct helicity dependent  transverse photovoltage is visible. For a bias applied along the a-axis, a transverse photovoltage emerges for CP illumination. Figure 3 (b) shows the amplitude of the helicity dependent transverse photovoltage $V^\mathrm{helical}$ vs bias voltage $V^\mathrm{bias}$ applied along the respective crystal directions. We find a finite $V^\mathrm{helical}$, which depends linearly on the longitudinal bias  if the bias is applied along the a-axis (yellow). In contrast, we detect no significant $V^\mathrm{helical}$ when the bias is applied along the b-axis (grey).\newline 
	In a next step, we perform time-resolved auto-correlation measurements to extract the characteristic time scales of the involved processes. 
	\begin{figure}
		\includegraphics{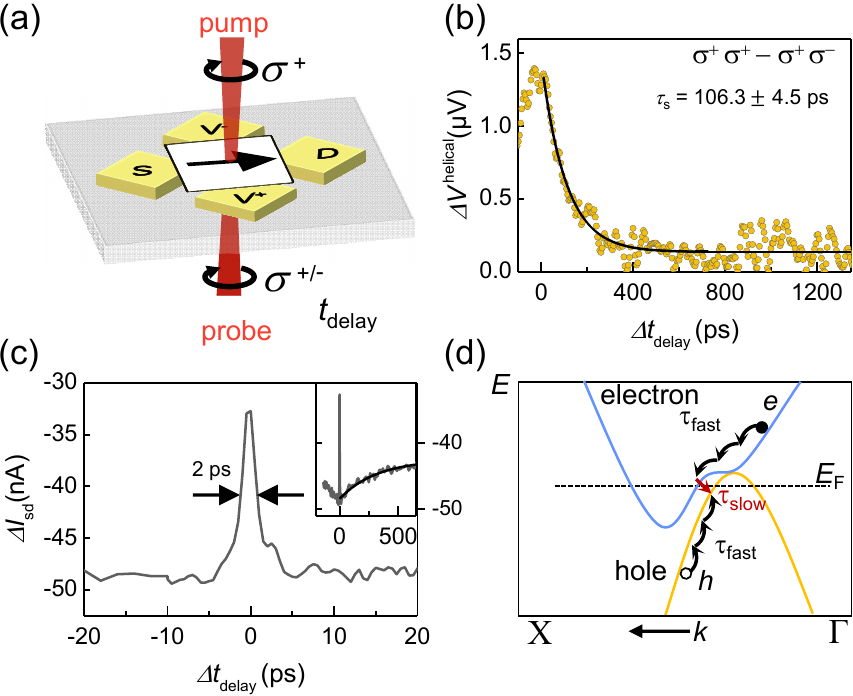}
		\caption{Time-resolved optoelectronic auto-correlation of spin and charge dynamics. (a) Schematic illustrations for measuring the auto-correlation of the helicity dependent transverse photovoltage. (b) Difference between co-polarized and cross-polarized probe laser excitation as a function of time delay. (c) Time-resolved pump/probe auto-correlation measurements of the longitudinal source-drain current with applied bias. The inset displays the same measurement on a longer time-scale. (d) Schematic band structure of WTe$_2$ near the Fermi level ($E_\mathrm{F}$) along the $\Gamma$ -X direction. The fast relaxation time $\tau_\mathrm{fast}$ describes the electron-phonon thermalization process and $\tau_\mathrm{slow}$ represents the phonon-assisted interband e-h recombination process.}
	\end{figure}
	Figure 4 (a) shows a schematic of our measurement geometry. A 150 fs pump pulse at a photon energy of 1.5 eV excites the center of the sample from the top at a fixed laser helicity. A probe pulse with identical duration and energy but variable helicity is delayed via a mechanical delay-stage and focused onto the same spot from the backside. Both pump and probe lasers are modulated at different frequencies, and the auto-correlation signal of the transverse photovoltage is detected at the sum-frequency with a lock-in amplifier. Figure 4 (b) shows the time-resolved difference of the transverse photovoltage for co-polarized and cross-polarized pump/probe excitation. We observe a finite difference directly after the excitation, which decays exponentially on a timescale of $\tau_\mathrm{slow} \approx 106.3 \pm 4.5$ ps. We interpret the difference between co-polarized and cross-polarized auto-correlation of the transverse photovoltage as a signature of the photo-excited spin population. Intriguingly, the decay time surpasses the electron-phonon relaxation time of a few ps by orders of magnitude \cite{Dai2015,Caputo2018}. In fact, the slow time-scale $\tau_\mathrm{slow}$ of spin relaxation was reported to be limited by the phonon-assisted recombination of momentum indirect electron-hole pairs, suggesting, that the charge carrier-relaxation to the Fermi-energy does not significantly randomize the spin-polarization \citep{Wang2018}. Figure 4 (c) shows the time-resolved auto-correlation of the longitudinal photo-current measured in the direction between the source and drain contacts. In contrast to the transverse photovoltage, the main time-scale dominating the longitudinal photocurrent, which we denote as $\tau_\mathrm{fast}$ is in the order of 2 ps and can be interpreted as an instantaneous photo-conductance increase, which is limited by the phonon-mediated charge carrier-relaxation to the Fermi-energy \cite{Wang2018,Dai2015,Caputo2018}. We note, that a slower time-scale is also detectable in the photo-current to a lesser extent. This indicates that the charge carrier spin does also impact the longitudinal charge transport, e.g. by anisotropic optical absorption as a consequence of a chiral anomaly in Weyl-semimetals \cite{Ashby2014,Mukherjee2017,Yang2015}. Figure 4(d)	shows a schematic energy band-diagram of the electron and hole pockets along the $\Gamma - X $ direction. The phonon-mediated carrier relaxation and recombination channels are indicated by their characteristic time-scales $\tau_\mathrm{fast}$ and $\tau_\mathrm{slow}$, respectively.
	\\
	\paragraph{Discussion}
	Our experiments demonstrate the observation of a photon helicity dependent transverse photovoltage in few-layer WTe$_2$ under an applied longitudinal bias. We interpret our findings to stem from an anomalous Hall-conductivity as a result of a helicity-induced  anisotropy in the system's spin-Hall conductivity $\sigma_H^s$ based on the following arguments. In the presence of an electric field, an electron with eigenenergy $\epsilon_n(\textbf{k})$ occupying a specific band $n$ can pick up an additional anomalous velocity contribution, which is proportional to the Berry curvature $\Omega_n(\textbf{k})$ of the specific band \cite{Xiao2010}. The velocity reads \cite{Xiao2010}
	\begin{equation}
	v_n(\textbf{k})=\frac{\delta\epsilon_n(\textbf{k})}{\hbar\delta\textbf{k}}-\frac{e}{\hbar}\textbf{E}\times \Omega_n(\textbf{k}),
	\end{equation}
	where the second term is the anomalous velocity contribution. It is always transverse to the electric field and is responsible for various Hall effects \cite{Xiao2010}. The Berry curvature $\Omega_n(\textbf{k})$ is the curl of the phase shift of a wave-function between two points in $k$-space and it is directly related to the symmetry of the system's Hamiltonian. An electron's velocity must be unchanged undergoing symmetry operations that reflect the symmetries of the unperturbed system \cite{Xiao2010}. Accordingly, eq. 1 dictates that $\Omega_n(\textbf{k})=-\Omega_n(-\textbf{k})$ if the system is invariant under time-reversal and $\Omega_n(\textbf{k})=\Omega_n(-\textbf{k})$ if the system is invariant under spatial inversion \cite{Xiao2010}. One can directly see that the Berry-curvature must vanish in systems, which have both inversion- and time-reversal symmetry. If time-reversal symmetry is broken, a net $\Omega_n$ can occur when integrating across the Brillouin-zone leading to a net anomalous Hall conductivity and a corresponding Hall voltage under applied bias. If inversion symmetry is broken, time-reversal dictates, that the net $\Omega_n$ is zero, when integrating across the Brillouin-zone, but it can exhibit a dipolar structure where opposite points in $k$-space exhibit opposite $\Omega_n(\textbf{k})$ and opposite anomalous velocities \cite{Xiao2010,Shi2018a,Zhang2018c}.  At an additional presence of spin-orbit coupling, a lifted spin-degeneracy at opposite points in $k$-space can lead to a net $\sigma_H^s$ and a transverse spin-transport perpendicular to the electric field. Indeed, the breaking of inversion-symmetry in WTe$_2$ along one crystal axis is predicted to induce Rashba and Zeeman type spin-orbit coupling as well as a dipolar Berry curvature \cite{Shi2018a}. Such a Berry curvature dipole is consistent with the crystal axis dependence of our measured transverse photovoltage (Fig. 3 and S3). We find a bias dependent transverse photovoltage $V^\mathrm{helical}$ only if the bias is applied along the a-axis orthogonal to the mirror-plane M$_\mathrm{a}$. The bias-dependence clearly demonstrates that the observed $V^\mathrm{helical}$ is not caused by a local photo-voltage generation e.g. due to built-in fields or impurities, but it is intrinsic in nature and proportional to the applied external electric field \cite{Seifert2018}. The transverse photovoltage switches polarity with the helicity of the exciting laser, which establishes that the underlying origin of the transverse conductivity is either directly based on the photon chirality \cite{Chan2016} or on the corresponding excited spin-density in the WTe$_2$ \cite{Seifert2018,Lee2018,Liu2018}. As we detect the $V^\mathrm{helical}$ for a 150 fs pulsed laser excitation on a long characteristic time scale $\approx 100$ ps, we identify the transverse photovoltage to stem from the excited spin-density on the non-equilibrium Fermi surface, rather than from a laser field driven anomalous Hall conductivity as suggested for a CW-laser excitation \cite{Chan2016}. In both cases, however, the underlying Hamiltonian is not invariant under time reversal, and in turn, it allows for a net anomalous Hall conductivity as well as a finite Hall voltage under applied bias. In principle, charge carriers can also acquire an anomalous velocity from extrinsic mechanisms such as skew-scattering and side-jump contributions, which can also be proportional to the Berry-curvature \cite{Sinitsyn2005,Onoda2006}. Experimentally, we generate the signal locally within the laser spot ($\approx 1.5$ \textmu m) and detect the photovoltage at rather macroscopic distances of $~14$ \textmu m. In metallic systems such as WTe$_2$ the detection occurs instantaneously according to the so-called Shockley-Ramo theorem \cite{Song2014a}. Therefore, we cannot distinguish between intrinsic Berry-phase effects and extrinsic drift or diffusive contributions. However, the linear dependence of our observed $V^\mathrm{helical}$ on the applied bias makes a side-jump contribuion less likely, as the latter is expected to show a non-linear contribution in the presence of non-zero Berry-curvature \cite{Onoda2006,Nagaosa2010}. Rationalized by this microscopic model, we interpret the transverse conductivity to stem from a laser-helicity induced transport anisotropy of an excited spin-ensemble as observed in topological insulators \cite{Seifert2018,Lee2018,Liu2018}. Without illumination, the bipolar distribution of the Berry-curvature gives rise to a spin-transport perpendicular to the applied bias. As time-reversal symmetry is preserved, the anomalous velocities of opposite spin orientations must cancel out. No net Hall-voltage can be observed. 
	\paragraph*{Conclusion}
	We explore the transverse conductivity in few layer WTe$_2$ under illumination and electrical current flow using polarization and time-resolved transverse photovoltage measurements. Our findings suggest that a helicity induced symmetry-breaking of an otherwise isotropic spin-current gives rise to an anomalous Hall-voltage. We further reveal, that the characteristic decay-time of the anomalous Hall-voltage exceeds the carrier-phonon scattering time by orders of magnitude but is consistent with the spin-lifetime of carriers in the momentum-indirect electron and hole pockets in WTe$_2$. 
	
	\paragraph{Acknowledgements}
	We thank T. Schmidt and M. Burghard for discussions. This work was supported by the DFG via SPP 1666 (grant HO 3324/8), the Center of NanoScience (CeNS) in Munich, and the Munich Quantum Center (MQC). C. K. acknowledges funding by the Molecular Foundry supported by the Office of Science, Office of Basic Energy Sciences, of the US Department of Energy under Contract No. DE-AC02-05CH11231.
	
	\section{References}
	
	\bibliography{library}

\clearpage
\newpage

\beginsupplement

	\section{Supplementary information \newline Photo-induced anomalous Hall effect in the type-II Weyl-semimetal WTe$_2$ at room-temperature 
	}
	\begin{figure}[h]
		\includegraphics[scale=1]{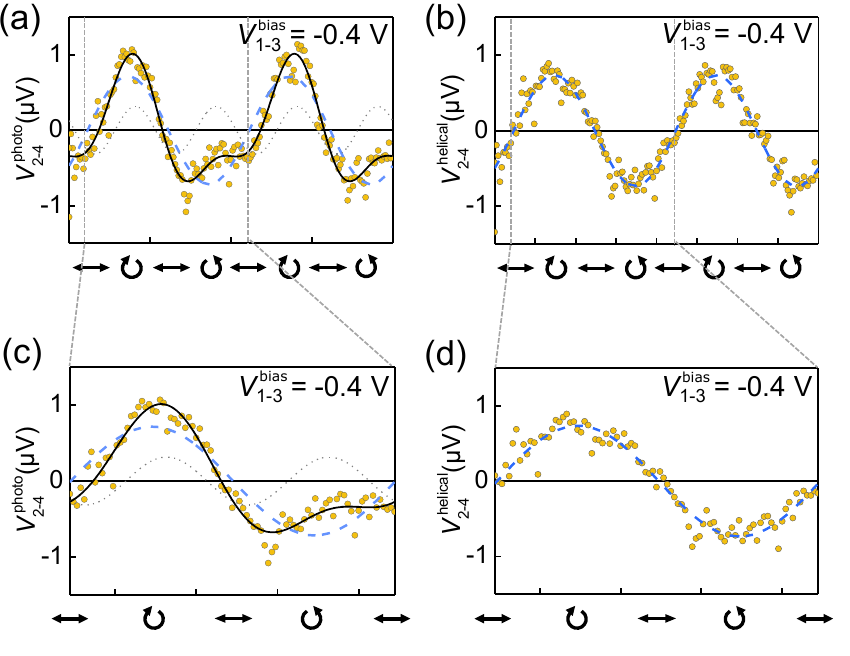}
		\caption{Fit procedure for polarization-resolved transverse photovoltage measurements. (a), Polarization dependence of the transverse photovoltage $V^\mathrm{photo}_{2-4}$, for applied bias voltages of $V^\mathrm{bias}_{1-3}$ = - 0.4 V as in Fig. 2 (b) in the main manuscript. The curve is fitted with an oscillation (black solid line) containing both sin$(2\alpha)$ (blue dashed line) and sin$(4\alpha)$ (grey dotted line) components for the helicity and linear polarization contributions where $\alpha$ is the rotation angle of the quarter waveplate. (b), Helicity dependent transverse photovoltage $V^\mathrm{helical}_{2-4}$ where the modulation with the linear polarization sin$(4\alpha)$ is subtracted from the data. (c) and (d), Corresponding zoom in on the curves in (a) and (b) as shown in Fig. 2 (e) of the main manuscript (indicated by the grey dashed lines).
		}
	\end{figure}
	\begin{figure}
		\includegraphics[scale=1]{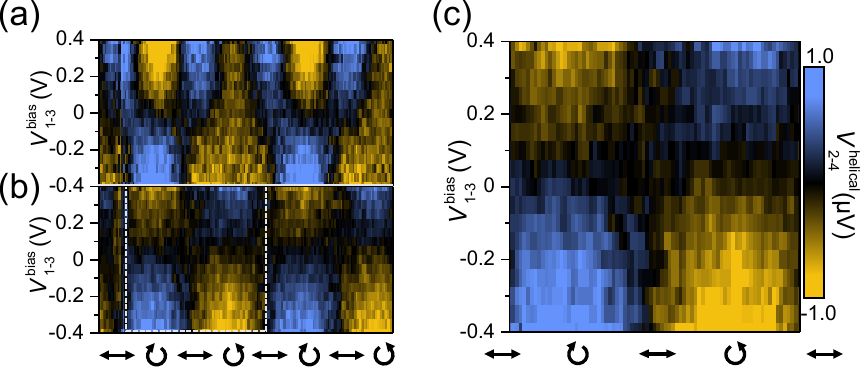}
		\caption{Fit procedure for voltage and polarization-resolved transverse photovoltage measurements. (a), Bias $V^\mathrm{bias}_{1-3}$ dependence of the polarization dependent transverse photovoltage $V^\mathrm{photo}_{2-4}$ as a function of laser polarization. (b), Bias $V^\mathrm{bias}_{1-3}$ dependence of the helicity dependent transverse photovoltage $V^\mathrm{helical}_{2-4}$ as in (a) with the linear polarization contribution subtracted as in supplementary figure 1 for each bias voltage $V^\mathrm{bias}_{1-3}$. (c), Corresponding zoom in on (b) as shown in Fig. 2 (d) of the main manuscript (indicated by the white dashed lines)
		}
	\end{figure}
	
	\begin{figure}
		\includegraphics[scale=1]{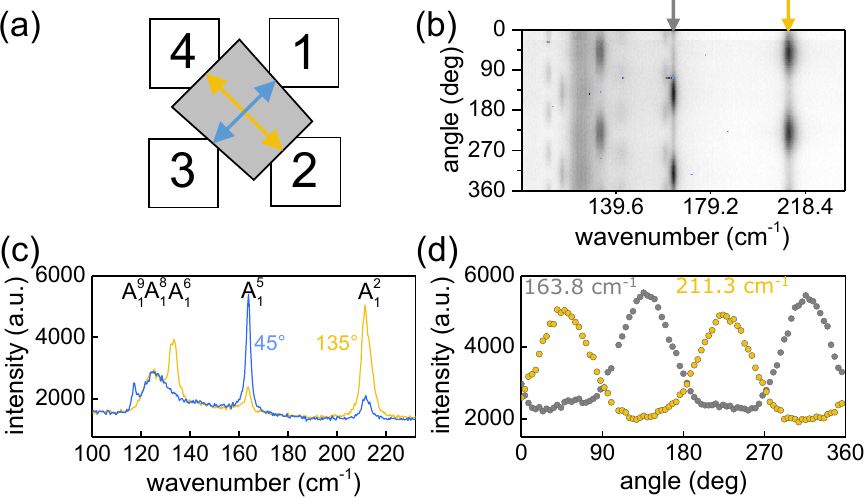}
		\caption{Raman signature of crystal axis configurations. (a), Sample geometry and contact configuration. The blue and yellow arrows indicate laser polarization axes parallel to the measurement configurations is Fig. 3 of the main manuscript. (b), Polarization resolved Raman spectra of the WTe$_2$ crystal for different linear excitation polarizations. The polarization is rotated with a half-waveplate. (c), Raman spectra for the laser polarizations indicated by the blue and yellow arrow in (a). (d), Cuts through the polarization resolved Raman spectra in (b) along the wavenumbers 163.8 cm$^{-1}$ and 211.3 cm$^{-1}$ as indicated by the grey and yellow arrows in (b). The maxima of the Raman modes at 163.8 cm$^{-1}$ and 211.3 cm$^{-1}$ indicate a laser polarization along the b-axis and a-axis respectively \cite{Song2016}. }
	\end{figure}

\end{document}